\documentclass[11pt]{article}

\usepackage[left=2cm, right=2cm, top=2.5cm, bottom=2.5cm]{geometry}
\usepackage[x11names]{xcolor}
\usepackage{fancyhdr, amssymb, cancel, amsmath, graphicx, pgfplots, tikz}

\newcommand{\stylecolor}{black}

\usepackage[colorlinks=true, urlcolor=violet, linkcolor=blue, citecolor=red, hyperindex=true, linktocpage=true]{hyperref}

\usepackage[explicit]{titlesec}

\newcommand*\sectionlabel{}
\titleformat{\section}
  {\gdef\sectionlabel{}
   \Large\bfseries\scshape}
  {\gdef\sectionlabel{\thesection. }}{0pt}
  {\begin{tikzpicture}[remember picture,overlay]
	\draw (-0.2, 0) node[right] {\textsf{\sectionlabel#1}};
	\draw[thick] (0, -0.4) -- (\textwidth, -0.4);
       \end{tikzpicture}
  }
\titlespacing*{\section}{0pt}{15pt}{20pt}

\newcommand*\subsectionlabel{}
\titleformat{\subsection}
  {\gdef\subsectionlabel{}
   \large\bfseries\scshape}
  {\gdef\subsectionlabel{\thesubsection.\ \  }}{0pt}
  {\begin{tikzpicture}[remember picture,overlay]
    	\draw (-0.15, 0) node[right] {\textsf{\subsectionlabel#1}};
       \end{tikzpicture}
  }
\titlespacing*{\subsection}{0pt}{10pt}{10pt}

\pgfplotsset{every axis legend/.append style={at={(1.02,1)},anchor=north west}}

\newcommand{\titletext}{Parallel implementation of fast randomized algorithms for the decomposition of low rank matrices}

\begin{document}

\pagestyle{fancy}
\renewcommand{\headrulewidth}{0pt}
\fancyhead{}

\fancyfoot{}
\fancyfoot[C] {\textsf{\textbf{\thepage}}}

\begin{equation*}
\begin{tikzpicture}
\draw (0.5\textwidth, -3) node[text width = \textwidth] {{\huge \begin{center} \color{\stylecolor} \textsf{\textbf{\titletext}} \end{center}}}; 
\end{tikzpicture}
\end{equation*}
\begin{equation*}
\begin{tikzpicture}
\draw (0.5\textwidth, 0.1) node[text width=\textwidth] {\large \color{black} \textsf{Andrew Lucas$^{a,b}$, Mark Stalzer$^b$ and John Feo$^c$}};
\draw (0.5\textwidth, -0.5) node[text width=\textwidth] {\small $\;^a$\textsf{Department of Physics, Stanford University, Stanford, CA 94305, USA}};
\draw (0.5\textwidth, -1) node[text width=\textwidth] {\small $\;^b$\textsf{Center for Advanced Computing Research, California Institute of Technology, Pasadena, CA 91125, USA}};
\draw (0.5\textwidth, -1.5) node[text width=\textwidth] {\small $\;^c$\textsf{Center for Adaptive Supercomputing Software, Pacific Northwest National Laboratory, Richland, WA 99352, USA}};
\end{tikzpicture}
\end{equation*}
\begin{equation*}
\begin{tikzpicture}
\draw (0.5\textwidth, -6) node[below, text width=0.8\textwidth] {\small  We analyze the parallel performance of randomized interpolative decomposition by decomposing low rank complex-valued Gaussian random matrices up to 64 GB.  We chose a Cray XMT supercomputer as it provides an almost ideal PRAM model permitting quick investigation of parallel algorithms without obfuscation from hardware idiosyncrasies. We obtain that on non-square matrices performance becomes very good, with overall runtime over 70 times faster on 128 processors.  We also verify that numerically discovered error bounds still hold on matrices nearly two orders of magnitude larger than those previously tested. };  
\end{tikzpicture}
\end{equation*}
\begin{equation*}
\begin{tikzpicture}
\draw (0, -13.1) node[right] {\texttt{stalzer at caltech.edu}};
\draw (\textwidth, -13.1) node[left] {\textsf{\today}};
\end{tikzpicture}
\end{equation*}

\tableofcontents

\section{Introduction}
Computational mathematics and science increasingly involves calculations using very large matrices which may easily be hundreds of gigabytes in size.  Nonetheless, many fundamental matrix algorithms such as QR factorization are not only very slow, with flop complexities of $\mathrm{O}(n^3)$ common, but also scale poorly on parallel machines due to frequent global communications.

A few years ago, it was proposed in a series of papers \cite{rokyale1, rokyale2, rokpnas} that a procedure, coined interpolative decomposition (ID), could decompose an $m\times n$ matrix, with approximate rank $k$, into two much smaller matrices.  Furthermore, when using a probabilistic algorithm to perform the ID, one had quadratic scaling behavior and much of the algorithm could be naturally parallelized.   Performing an ID on a large low-rank matrix not only allows for it to be stored in a much smaller amount of memory, but it allows for many core operations (such as matrix multiplication) to run significantly faster.  Furthermore, the ID and similar randomized algorithms can serve as the basis for fast methods for the singular value decomposition (SVD) \cite{rokpnas} and principal component analysis \cite{rok1}, as well as tensor skeleton generalizations \cite{drineas} and least squares problems \cite{boutsidis, rok2, rudelson}; for a much more thorough list see \cite{martinsson}.

Although there is increasing interest in probabilistic algorithms, we are not aware of these algorithms having been tested on extremely large matrices where massive parallelization is essential.  In this paper, we present results obtained on a Cray XMT supercomputer by performing a randomized ID on matrices over 100 GB in size, which is two orders of magnitude larger than previous matrices studied (without making calls out of RAM).   In doing so, we confirm the conjecture that such algorithms can run efficiently in parallel, demonstrating that for many matrices, the algorithm's parallelism scales by at least two orders of magnitude.  We also confirm numerical results which suggested that the error bounds on the procedure grow slowly with matrix size, implying that these procedures are appropriate for high-precision computation on extremely large matrices.

\section{The Randomized Interpolative Decomposition Algorithm}
Let $A$ be an $m\times n$ complex matrix with approximate rank $k$, with $k\ll m,n$.  By approximate rank, we mean that the $(k+1)^{\mathrm{st}}$ largest singular value, $\sigma_{k+1}$, is small. The goal of the randomized ID algorithm \cite{rokpnas} is to find an $m\times k $ matrix $B$ and a $k \times n$ matrix $P$ such that \begin{equation}
A\approx BP.
\end{equation}It is well-known from the Eckart-Young theorem \cite{eckart} that the best possible choices of $B$ and $P$ come from performing a spectral value decomposition (SVD) \begin{equation}
A=U\Sigma V,
\end{equation} replacing $\Sigma$ with a truncated diagonal matrix $\Sigma_k$ with only the largest $k$ elements retained, and setting $B$ equal to the leftmost $k$ columns of $U$, and letting $P$ equal to the topmost $k$ rows of $\Sigma_k V$.  However, the SVD has runtime $\mathrm{O}(mn^2)$ and there are no parallel algorithms for the SVD which work very fast on matrices of arbitrary structure, with the exception of algorithms derived from the randomized ID.

The variations of the randomized ID provide an algorithm to solve the above problem which is efficient in parallel on a matrix with arbitrary structure, and a typical randomization scheme will work well on nearly every matrix.   In fact, we can be more precise.   Performing an asymptotic expansion of the results of Observation 21 of \cite{rokyale1} in the limit $m,n \gg k \gg 1$, with probability no smaller than $1-\epsilon$,\begin{equation}
\frac{\| A-BP \|_2}{\sigma_{k+1}} \le 50 \sqrt{mn} \left(\frac{1}{\epsilon}\right)^{\frac{1}{k}} \label{eq:errorbound}
\end{equation}
Note that the right hand side in the optimal case is 1.   Even for $\epsilon \sim 10^{-20}$, so long as $k \gtrsim 100$, the $\epsilon$-dependent factor is essentially negligible.  The algorithm we chose to implement is a modification of the one for which a particularly tight error bound has been derived, but in practice the error bounds for our algorithm have appeared similarly good \cite{rokyale2}, and we will confirm that this bound is obeyed in our numerical results in Section~\ref{sec:xmt}.

Let us now briefly describe the mathematics and structure behind the algorithm.    The intuitive way to think of the algorithm is that it compresses the matrix $A$ into something with very few rows by attempting to extract all information about the linear independence: i.e., to express the matrix in terms of a basis of vectors, each with $\mathrm{O}(k)$, as opposed to $\mathrm{O}(m,n)$, nontrivial elements.  From there, it performs highly accurate QR factorizations on a subset of columns: given an orthonormal basis for these reduced columns, the remainder of the matrix can be rapidly factored.  The key to the algorithm is that the only parts of the algorithm which are slow and not efficiently parallelized are only run on a very tiny matrix, compared to the size of $A$.

The first step of the algorithm corresponds to the randomization of the matrix $A$, which is achieved by compressing it into an $l\times n$ matrix $Y$.  The parameter $l$ can be chosen by the user and it will correspond to how many rows of the randomized matrix we will extract.  Clearly we must choose $l\ge k$ in order to ensure that there can be $k$ linearly independent rows/columns.   We always chose $l=2k$ to allow for the possibility that some rows were linearly dependent while still keeping the randomized matrix small, and in practice this choice was always adequate, and also allowed us to estimate the error bound (\ref{eq:errorbound}).  The compression is done by writing \begin{equation}
Y=SFDA
\end{equation}
where $S$ is a $l\times m$ matrix with entries
\begin{equation}
S_{jk} = \delta_{js_k}
\end{equation}with $\delta_{ij}$ the Kronecker $\delta$ (1 when its indices are equal, otherwise 0) and $\lbrace s_1,\ldots, s_l\rbrace$ are i.i.d.\footnote{i.i.d. = independent and identically distributed} random variables uniformly distributed on $\lbrace 1,\ldots,m\rbrace$, $F$ is the $m\times m$ fast Fourier transform (FFT) operator:\begin{equation}
F_{jk} = \mathrm{e}^{-2\pi\mathrm{i}(j-1)(k-1)/m},
\end{equation} and $D$ is a diagonal $m\times m$ matrix of random complex phases: \begin{equation}
D_{jk} = \mathrm{e}^{2\pi \mathrm{i}\phi_j} \delta_{jk} ,
\end{equation}
with $\lbrace \phi_1,\ldots, \phi_m\rbrace$ are i.i.d. uniform random variables on $[0,1]$.    In short, the matrix $D$ multiplies each row by a random complex phase; the matrix $F$ performs a FFT on each column, and the matrix $S$ simply sets $Y$ equal to a matrix consisting of $l$ randomly chosen rows from the Fourier transformed matrix $FDA$.   We cannot find a generic class of matrices for which this randomization procedure should fail (i.e., $\mathrm{rank}(Y) < k$) with high probability, although even if this event should occur, repeating the algorithm with a different instance of the random matrices $S$ and $D$ should result in a new choice of $Y$ with sufficient rank.

The next step consists of a highly approximate QR factorization on the matrix $Y$: i.e., expressing \begin{equation}
Y\approx QR
\end{equation}where  $Q$ is an $l\times k$ matrix with orthonormal columns and $R$ is a $k\times n$ matrix of the form \begin{equation}
R = \left(\begin{array}{cc} R_1 &\ R_2\end{array}\right)
\end{equation} where $R_1$ is a $k\times k$ upper triangular matrix, and $R_2$ is a $k\times(n-k)$ matrix \cite{cheng}.   As with typical QR factorizations, the aim here is to find an orthonormal set of vectors to serve as an approximate basis for the columns of $Y$.    We should also note that, in general, it may be the case that the first $k$ columns of the matrix do not contain (with high probability) the $k$ linearly independent vectors of highest weight.  If this is the case, then we must first multiply $A$  by an appropriate permutation matrix, before randomizing $A$ to $Y$, so that the first $k$ columns are linearly independent and contain the $k$ most weighted vectors.

The final step consists of forming $B$ and $P$.  $B$ is found by taking the left-most $m\times k$ matrix of $A$.  The matrix $P$ is constructed as follows: first, find the matrix $T$ such that \begin{equation}
R_2 = R_1T.
\end{equation}This problem can be solved exactly because $R_1$ is upper triangular, and it reduces to the problem of solving $L\mathbf{v} = \mathbf{w}$ for triangular $L$ and unknown $\mathbf{v}$, given $\mathbf{w}$ \cite{golub}.   From here, one sets \begin{equation}
P = \left(\begin{array}{cc} I &\ T \end{array}\right)
\end{equation}
where $I$ is the $k\times k$ identity matrix.  In practice, we combined the QR factorization of $R_2$ with the factorization of $R_2=R_1T$, as this process can be done simultaneously on all columns.   In our benchmarking, we have referred to this last phase as the factorization of $R$, for simplicity.

Overall, the complexity of the algorithm is given by $\mathrm{O}(mn\log m + lk^2 + k(l+k)(n-k))$, with the terms corresponding to the FFT, QR factorization of $R_1$ and the combined factorization of $R_2$ respectively.  

It is worth noting that there are alternative randomization algorithms, but that the algorithm proposed here is likely the most generically effective as it works well independent of the structure of the matrix $A$, and the slowest step is the FFT, for which there are many excellent implementations.  However, if a faster method of computing the randomization step is available, that should be used instead (see, e.g., \cite{rokyale1, rokpnas}).

\section{Implementation and Results}
\label{sec:xmt}
\subsection{Details of the Cray XMT}
The XMT is a shared-memory, multithreaded machine.  There is no cache and no local memory; all processors can access all memory locations in the same time.  Each processor has 128 registers sets, 128 program counters (one per register set), and a single, three instruction wide execution pipeline.  The pipeline can execute one memory and three floating point operations per cycle.  Up to 128 different software threads can be co-scheduled per processor.  On each cycle, a processor chooses a software thread with a ready instruction and executes it.  As long as one of the 128 threads has a ready instruction the processor remains busy.  Thus, long latency operations such as memory accesses, synchronization operations, or runtime system calls are tolerated via parallelism.  The parallel performance and scalability of an algorithm is a function of only its parallelism.  The Cray XMT is almost an ideal PRAM system supporting equally a wide variety of parallel techniques including data and task parallelism, dataflow, and recursion.  In our study, we used data parallelism, collapsing loop nests to generate a sufficient number of independent threads to saturate a 128 processor Cray XMT (about 100 threads per processor). In many instances, the compiler collapsed the nests for us; but, where it did not we used pragmas to force the issue. 
\subsection{Randomized ID Implementation}
We now briefly describe the algorithm's intrinsic parallelism and our implementation.  Because the FFT can be performed on each column separately, and the factorization of $R$ can also be done individually for each column, these parts of the algorithm exhibit both coarse and fine-grain parallelism and thus we found them to scale decently.

Ultimately, the clear bottleneck in this procedure is the QR factorization, which must be done extremely accurately.   The typical algorithm to use here is a Gram-Schmidt algorithm.   We chose to use a classical GS algorithm with iteration -- this is the most numerically stable variant of GS \cite{bjorck}, and it also works well in highly parallel contexts \cite{lingen}, beating out an iterated modified GS \cite{hoffman}.    After the completion of this work, we learned that performing the QR factorization with Householder reflections should result in similar stability with only half the runtime.

Due to the specific nature of the machine on which we implemented this algorithm, we had to write custom implementations for all of our code with the sole exception of very rudimentary probabilistic functions.   We implemented the standard radix-4 Cooley-Tukey algorithm as an outer iterative loop over stages and a parallel nested loop over butterflies.     The factorization of $R$ proceeded column-wise in parallel, with each processor allowed to work on a separate column.    
\subsection{Benchmarking Results}
The algorithm was implemented in C with 64-bit precision floating point arithmetic.  The matrix $A$ was formed by constructing $B$ and $P$ to be Gaussian random matrices with complex entries and the appropriate dimensions, and setting $A=BP$.   Due to precision error, we should expect that $\sigma_{k+1} \gtrsim \sqrt{2\min(m,n)}\times \delta$, where $\delta\sim 10^{-16}$ is the precision error of the multiplication $BP$ (this estimate is rigorous for a square Gaussian random matrix \cite{rokyale1}).   The reason we chose random low rank matrices to benchmark with is that these matrices have almost no exploitable structure, other than their rank.  Many fast algorithms for matrix factorization and decomposition rely on assumptions about convenient matrix structure, such as sparsity.   Our aim is to demonstrate the effectiveness of these algorithms regardless of any underlying structure.

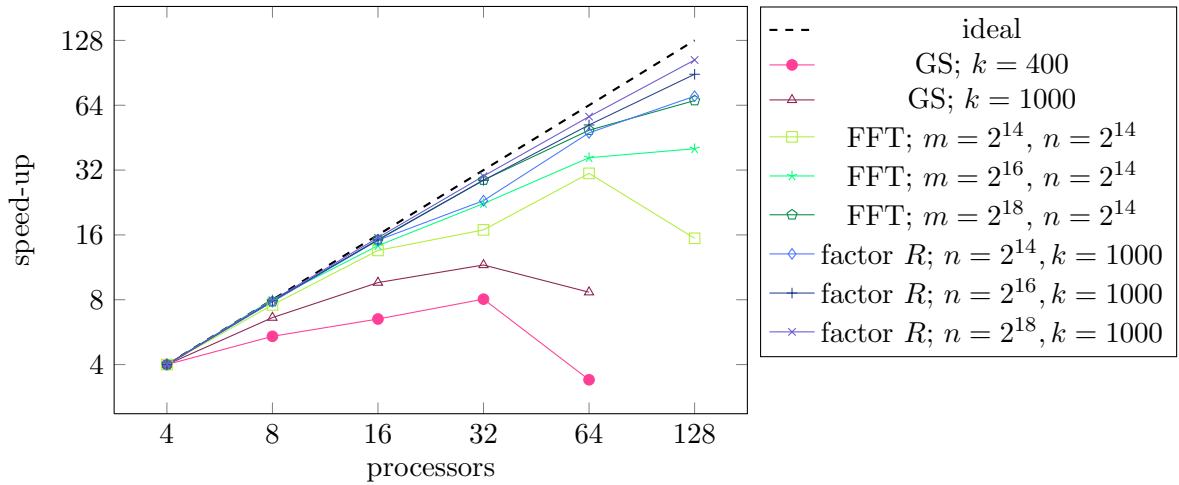
\begin{figure}
\centering
\begin{tikzpicture}
\begin{loglogaxis}[width=10cm, height=7cm, xlabel=processors, ylabel=speed-up,
xtick={4,8,16,32,64,128}, ytick={4,8,16,32,64,128}, xticklabels={4,8,16,32,64,128}, yticklabels={4,8,16,32,64,128}]
\addplot[color=black, thick, dashed] coordinates {(4,4) (128,128)};
\addlegendentry{ideal};

\addplot[color=VioletRed1, mark=*] coordinates {(4,4) (8, 5.41) (16, 6.51) (32, 8.05) (64, 3.4)};
\addlegendentry{GS; $k=400$};

\addplot[color=VioletRed4, mark=triangle] coordinates {(4,4) (8, 6.61) (16, 9.61) (32, 11.6) (64, 8.67)};
\addlegendentry{GS; $k=1000$};

\addplot[color=OliveDrab2, mark=square] coordinates {(4,4) (8, 7.57) (16, 13.55) (32, 16.88) (64, 30.89) (128, 15.43)};
\addlegendentry{FFT; $m=2^{14}$, $n=2^{14}$};

\addplot[color=SpringGreen1, mark=star] coordinates {(4,4) (8, 8.01) (16, 14.26) (32, 22.37) (64, 36.53) (128, 40.2)};
\addlegendentry{FFT; $m=2^{16}$, $n=2^{14}$};

\addplot[color=SpringGreen4, mark=pentagon] coordinates {(4,4) (8, 7.84) (16, 15.21) (32, 28.76) (64, 48.99) (128, 67.28)};
\addlegendentry{FFT; $m=2^{18}$, $n=2^{14}$};

\addplot[color=RoyalBlue1, mark=diamond] coordinates {(4,4) (8, 7.83) (16,15.12) (32, 23.08) (64, 47.63) (128, 70.49)};
\addlegendentry{factor $R$; $n=2^{14}, k=1000$};

\addplot[color=RoyalBlue4, mark=+] coordinates {(4, 4) (8, 7.91) (16, 15.15) (32, 28.69) (64, 51.84) (128, 89.11)};
\addlegendentry{factor $R$; $n=2^{16}, k=1000$};

\addplot[color=SlateBlue3, mark=x] coordinates {(4,4) (8, 7.96) (16, 15.53) (32, 30) (64, 56.66) (128, 103.8)};
\addlegendentry{factor $R$; $n=2^{18},  k=1000$};
\end{loglogaxis}
\end{tikzpicture}
\caption{Parallel speed-up of various processes on the XMT.  For all runs, take $l=2k$.}
\label{factorr}
\end{figure}

As expected theoretically, the FFT runtime was dominated by $m$, the GS runtime was dominated by $k$, and the $R$ factorization runtime was dominated by $n$.  Figure \ref{factorr} shows the parallel efficiency of the factorization of $R$ on an increasing number of processors.  Note that we are defining parallel efficiency in the standard way.
We saw parallel speed-ups of over 100 times faster on 128 processors for $n=2^{18}$ for the factorization of $R$.  In contrast, we saw gains of only about 70 times faster on 128 processes for the FFT.  As the factorization of our algorithm ran particularly fast, we can conclude that at least using our implementation, this algorithm runs most efficiently on fairly skinny matrices with $m<n$, although this may be machine specific.  Note that one can always arrange things easily so that $n\ge m$ by simply taking a transpose, which will not affect the rank.

\begin{figure}
\centering
\begin{tikzpicture}
\begin{loglogaxis}[width=10cm, height=7cm, xlabel=processors, ylabel=speed-up,
xtick={4,8,16,32,64,128}, ytick={4,8,16,32,64,128}, xticklabels={4,8,16,32,64,128}, yticklabels={4,8,16,32,64,128}]
\addplot[color=black, thick, dashed] coordinates {(4,4) (128,128)};
\addlegendentry{ideal};

\addplot[color=blue, mark=*] coordinates {(4,4) (8, 7.6) (16, 13.6)  (64, 32) (128, 16)};
\addlegendentry{$k=100$, $m=2^{14}$, $n=2^{14}$};

\addplot[color=DodgerBlue2, mark=star] coordinates {(4,4) (8, 8.04) (16, 14.26)  (32, 22.3) (64, 34.4) (128, 38)};
\addlegendentry{$k=100$, $m=2^{16}$, $n=2^{14}$};

\addplot[color=SpringGreen1, mark=triangle] coordinates {(4,4) (8, 7.66) (16, 14.8)  (32, 27.2) (64, 34.7) (128, 23)};
\addlegendentry{$k=400$, $m=2^{16}$, $n=2^{14}$};

\addplot[color=Green4, mark=square] coordinates {(4,4) (8, 7.78) (16, 15.14)  (32, 28) (64, 44.6) (128, 49)};
\addlegendentry{$k=400$, $m=2^{18}$, $n=2^{14}$};

\addplot[color=orange, mark=diamond] coordinates {(4,4) (8, 7.84) (16, 15)  (32, 23) (64, 38.18) (128, 44.5)};
\addlegendentry{$k=100$, $m=2^{16}$, $n=2^{16}$};

\addplot[color=brown, mark=pentagon] coordinates {(4,4) (8, 7.93) (16, 15.17)  (32, 28.1) (64, 50) (128, 52)};
\addlegendentry{$k=100$, $m=2^{16}$, $n=2^{16}$};

\addplot[color=Firebrick3, mark=x] coordinates {(4,4) (8, 7.7) (16, 15.15)  (32, 28.5) (64, 42) (128, 49)};
\addlegendentry{$k=400$, $m=2^{14}$, $n=2^{18}$};

\addplot[color=red, mark=+] coordinates {(4,4) (8, 7.86)  (32, 29) (64, 54.6) (128, 73.4)};
\addlegendentry{$k=1000$, $m=2^{14}$, $n=2^{18}$};

\end{loglogaxis}
\end{tikzpicture}
\caption{Parallel speed-up of the overall algorithm on the XMT.  For all runs, take $l=2k$.}
\label{ovef}
\end{figure}
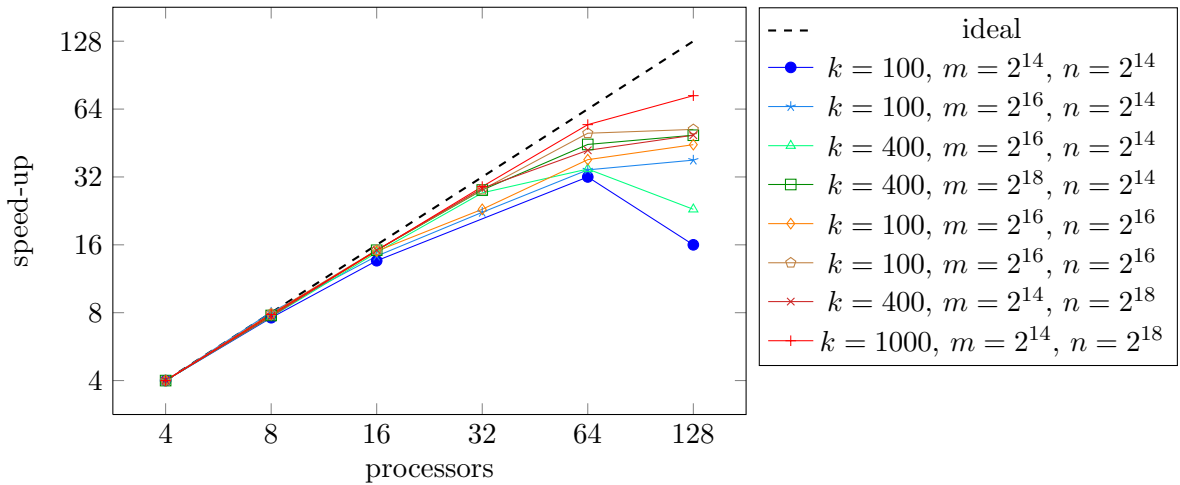

The overall speed-up of the algorithm is shown in Figure \ref{ovef}.   We saw excellent scaling by the time we reached $n=2^{18}$, except for on 128 processors where the FFT began performing much worse.  This did not happen for  $m>2^{14}$ and thus we expect that a more optimized FFT implementation would avoid this issue entirely.  Table 1 shows the overall runtime in seconds for the various processes on each of the runs shown in Figure \ref{ovef}; similarly, Table 2 gives the results for the Gram-Schmidt process, Table 3 for the FFT, and Table 4 for the factorization of $R$.    We did not run the code on either 1 or 2 processors due to constraints on our benchmarking time, and due to the excellent scaling of nearly all of the steps of the algorithm in between 4 and 8 processors.  

\begin{table}[here]
\centering
\begin{tabular}{|c|c|c|c|c|c|c|}\hline
matrix parameters &\ 4 &\ 8 &\ 16 &\ 32 &\ 64 &\ 128  \\ \hline
$k=100$, $m=2^{14}$, $n=2^{14}$ &\ 56.0 &\ 29.56  &\  16.61  &\  13.37 &\  7.16  &\ 14.04  \\ 
$k=100$, $m=2^{16}$, $n=2^{14}$ &\  188.6 &\ 95.2  &\ 53.52  &\ 34.1  &\ 21.88   &\  20.46  \\ 
$k=400$, $m=2^{16}$, $n=2^{14}$ &\ 381.1   &\  194.7 &\  103.3   &\  56.43   &\  44.2  &\  68.4  \\ 
$k=400$, $m=2^{18}$, $n=2^{14}$ &\ 923.9  &\  476.1 &\  244.7 &\ 133.6  &\  84.12  &\  76.83 \\ 
$k=100$, $m=2^{16}$, $n=2^{16}$ &\  732.8 &\  375.4 &\ 196.5  &\  129.4 &\  77.16   &\ 67.87  \\ 
$k=1000$, $m=2^{16}$, $n=2^{16}$ &\   5657 &\  2873 &\ 1492  &\  806.6 &\   455.3 &\  444.01 \\ 
$k=400$, $m=2^{14}$, $n=2^{18}$ &\ 3780  &\  1919 &\   1001 &\  531.4 &\  363.9   &\  310  \\ 
$k=1000$, $m=2^{14}$, $n=2^{18}$ &\  20167 &\ 10233 &\  -- &\  2894 &\    1479 &\  1099  \\  \hline
\end{tabular}
\caption{Runtime of the total algorithm, in seconds, for the given parameters and number of processors listed in the top row.}
\label{bigt}
\end{table}

\begin{table}[here]
\centering
\begin{tabular}{|c|c|c|c|c|c|c|}\hline
matrix parameters &\ 4 &\ 8 &\ 16 &\ 32 &\ 64 &\ 128  \\ \hline
$k=100$, $m=2^{14}$, $n=2^{14}$ &\ 43.0 &\ 22.7 &\ 12.69  &\  10.9 &\  5.57  &\   11.15 \\ 
$k=100$, $m=2^{16}$, $n=2^{14}$ &\  176.0 &\ 88.4 &\  49.63 &\  31.65 &\    19.4 &\  17.61 \\ 
$k=400$, $m=2^{16}$, $n=2^{14}$ &\  175.4  &\  88.4 &\  46.76  &\  25.85 &\   19.6 &\  17.70 \\ 
$k=400$, $m=2^{18}$, $n=2^{14}$ &\  718.0  &\ 363.4  &\  188.1  &\ 99.44   &\  60.86   &\ 42.23  \\ 
$k=100$, $m=2^{16}$, $n=2^{16}$ &\  684.3 &\  350.5 &\  183.3 &\  122.2 &\  72.39  &\  63.7  \\ 
$k=1000$, $m=2^{16}$, $n=2^{16}$ &\   687.8  &\  350.93 &\  183.82 &\ 114.9  &\  59.82   &\  64.71  \\ 
$k=400$, $m=2^{14}$, $n=2^{18}$ &\  683.6 &\ 360 &\  199.5  &\  114.3  &\  137.15  &\  148.1  \\ 
$k=1000$, $m=2^{14}$, $n=2^{18}$ &\  691  &\  362 &\  -- &\   123.9 &\   79.91 &\ 168  \\  \hline
\end{tabular}
\caption{Runtime of the FFT, in seconds, for the given parameters and number of processors listed in the top row.  Note that this is effectively independent of $k$.}
\label{bigt1}
\end{table}
\begin{table}[here]
\centering
\begin{tabular}{|c|c|c|c|c|c|c|}\hline
matrix parameters &\ 4 &\ 8 &\ 16 &\ 32 &\ 64 &\ 128  \\ \hline
$k=100$, $m=2^{14}$, $n=2^{14}$ &\ 0.23 &\ 0.36 &\  0.43 &\  0.53  &\  0.52   &\  2.11 \\ 
$k=100$, $m=2^{16}$, $n=2^{14}$ &\  0.23  &\  0.35 &\  0.44  &\  0.46 &\  1.41   &\  2.09  \\ 
$k=400$, $m=2^{16}$, $n=2^{14}$ &\  6.80 &\ 5.03 &\ 4.19  &\  3.38 &\   8.00 &\  39.3  \\ 
$k=400$, $m=2^{18}$, $n=2^{14}$ &\  6.86 &\  4.16 &\  4.12 &\   3.82 &\  6.66  &\ 23.2  \\ 
$k=100$, $m=2^{16}$, $n=2^{16}$ &\  0.22 &\ 0.24  &\ 0.39   &\  0.52 &\  1.04   &\ 2.0  \\ 
$k=1000$, $m=2^{16}$, $n=2^{16}$ &\  83.2 &\ 50.11  &\  34.65 &\  28.7 &\ 38.4   &\  175.9 \\ 
$k=400$, $m=2^{14}$, $n=2^{18}$ &\  6.66 &\  4.88 &\ 3.74  &\  3.61  &\    7.19 &\  42.74  \\ 
$k=1000$, $m=2^{14}$, $n=2^{18}$ &\ 82.02  &\  49.69 &\  --  &\  30.8 &\   29.32 &\  183.9 \\  \hline
\end{tabular}
\caption{Runtime of the Gram-Schmidt process, in seconds, for the given parameters and number of processors listed in the top row.   Note that it is effectively independent of $m$ and $n$.}
\label{bigt2}
\end{table}
\begin{table}[here]
\centering
\begin{tabular}{|c|c|c|c|c|c|c|}\hline
matrix parameters &\ 4 &\ 8 &\ 16 &\ 32 &\ 64 &\ 128  \\ \hline
$k=100$, $m=2^{14}$, $n=2^{14}$ &\ 12.5 &\ 6.50 &\  3.49  &\  1.93 &\   1.07 &\  0.78 \\ 
$k=100$, $m=2^{16}$, $n=2^{14}$ &\  12.4 &\ 6.47  &\  3.45  &\ 1.94   &\  1.07   &\  0.76  \\ 
$k=400$, $m=2^{16}$, $n=2^{14}$ &\  198.9  &\ 101.3  &\   52.3 &\  27.2 &\  16.6  &\  11.4 \\ 
$k=400$, $m=2^{18}$, $n=2^{14}$ &\  199.0 &\  108.5 &\  52.52   &\ 30.3  &\   16.6 &\  11.4 \\ 
$k=100$, $m=2^{16}$, $n=2^{16}$ &\  48.3 &\  24.65 &\  12.78  &\   6.74 &\   3.73 &\  2.17 \\ 
$k=1000$, $m=2^{16}$, $n=2^{16}$ &\  4886 &\ 2463 &\  1274  &\  663  &\  357.1  &\  203.4 \\ 
$k=400$, $m=2^{14}$, $n=2^{18}$ &\ 3090  &\ 1554  &\ 797.3  &\ 413.5  &\   219.5 &\  119.2 \\ 
$k=1000$, $m=2^{14}$, $n=2^{18}$ &\  19394  &\ 9821  &\  --  &\ 2739   &\ 1370   &\  747   \\  \hline
\end{tabular}
\caption{Runtime of the factorization of $R$ and $P$, in seconds, for the given parameters and number of processors listed in the top row.  Note that this is effectively independent of $m$.}
\label{bigt3}
\end{table}

As a final note, it is worthwhile to emphasize that as our benchmarking spanned an order of magnitude in matrix size, the increase in error on average was similarly an magnitude: from about $\sim 4\times 10^{-11}$ to $10^{-9}$.  Thus, we have found that the numerically determined low bounds on the error in this algorithm continue to hold on much larger matrices (recall that the proof of the error bound relied on a slightly different randomization scheme).  Table \ref{ertab} shows the errors bounds for some sample runs.  We found, as in \cite{rokpnas}, that the numerical error bound is satisfied, although reasonably tightly.  
\begin{table}[here]
\centering
\begin{tabular}{|c|c|}\hline
matrix parameters &\ $\| A-BP\|_2$ \\ \hline
$k=100$, $m=2^{14}$, $n=2^{14}$ &\ $5\times 10^{-11}$ \\
$k=100$, $m=2^{16}$, $n=2^{14}$ &\ $1\times 10^{-10}$ \\
$k=400$, $m=2^{16}$, $n=2^{14}$ &\ $2\times 10^{-10}$ \\
$k=400$, $m=2^{18}$, $n=2^{14}$ &\ $4\times 10^{-10}$ \\
$k=100$, $m=2^{16}$, $n=2^{16}$ &\ $2\times 10^{-10}$ \\
$k=1000$, $m=2^{16}$, $n=2^{16}$ &\ $6\times 10^{-10}$ \\
$k=400$, $m=2^{14}$, $n=2^{18}$ &\ $3\times 10^{-10}$ \\
$k=1000$, $m=2^{14}$, $n=2^{18}$ &\ $6\times 10^{-10}$ \\ \hline
\end{tabular}
\caption{The error, $\| A-BP\|_2$, for a variety of runs on the XMT.  Note that we would expect $\sigma_{k+1} \sim 10^{-14}$ to be of the order $10^{-14}$, and thus $\| A-BP\|_2 \sim 10^{-8}$ if our bound was tight.  For the larger matrices these spectral norm bound increases by a  factor of 8.}
\label{ertab}
\end{table}

As mentioned in the introduction, we are not aware that randomized algorithms have been tested on such large matrices using highly parallel implementations to date.   \cite{konda} presents an algorithm for a fast parallel SVD on bidiagonal matrices which obtains similar speed-ups to ours, although our algorithm works on any low rank matrix.

\section{Conclusion}
This letter demonstrates that the parallel implementations of randomized algorithms are efficient and scale well for very large matrices.  These algorithms do tend to become less ideal on 128 processors because the FFT is starved, but we emphasize that as our results can be obtained for \emph{arbitrary} matrices with low rank, even the scaling we found is very impressive.  

Furthermore, concerns about the breakdown of the randomized algorithm on large matrices approaching are unfounded -- we have demonstrated that the error in the algorithms slowly enough with matrix size that they should be adequate for many applications.  Finally, we stress that our benchmarking was done on dense random matrices -- it is quite likely that the scaling would be far better were the low rank matrices in question of a more specialized form where certain steps in the algorithm could be optimized further.

Optimizing the Gram-Schmidt and FFT algorithms on highly parallel architectures is an active area of research and it is almost assured that specialized code would be  faster than ours on these components.  However we did not feel it appropriate to emphasize the optimization of these algorithms in the preparation of this letter, as our focus was instead on the parallelization of randomized algorithms.  Work on improving these components of the algorithm, as well as finding specialized random algorithms for certain classes of large (low-rank) matrices is a worthwhile direction for future research.

\section*{Acknowledgements}
\addcontentsline{toc}{section}{Acknowledgements}
The authors would like to thank the staff at CACR for help with initial benchmarking of the algorithm and for advice on optimization, as well as the staff at PNNL for assistance and usage of the Cray XMT.   We would also like to thank V. Rokhlin for helpful comments.

\bibliographystyle{unsrt}
\addcontentsline{toc}{section}{References}
\bibliography{xmtmatrixbib}

\end{document}